# PERFORMANCE OF CACHE MEMORY SUBSYSTEMS FOR MULTICORE ARCHITECTURES

N. Ramasubramanian[1], Srinivas V.V.[2] and N. Ammasai Gounden[3]

[1, 2]Department of Computer Science and Engineering,
National Institute of Technology, Tiruchirappalli.
[3]Department of Electrical and Electronics Engineering,
National Institute of Technology, Tiruchirappalli.
`nrs@nitt.edu`
`206110022@nitt.edu`
`ammas@nitt.edu`

## ABSTRACT

*Advancements in multi-core have created interest among many research groups in finding out ways to harness the true power of processor cores. Recent research suggests that on-board component such as cache memory plays a crucial role in deciding the performance of multi-core systems. In this paper, performance of cache memory is evaluated through the parameters such as cache access time, miss rate and miss penalty. The influence of cache parameters over execution time is also discussed. Results obtained from simulated studies of multi-core environments with different instruction set architectures (ISA) like ALPHA and X86 are produced.*

## 1. INTRODUCTION

One of the important factors that influence execution time of a program is the cache access time [1]. Cache memory provides a quicker supply of data for execution by forming a bridge between the faster processor unit on one side and the relatively slower memory unit on the other side. While it is well known that cache memory helps in faster access of data, there is considerable interest among research groups to understand the impact of cache performance on the execution time for obtaining better performance of multi-core platforms [2].

Latest advancements in cache memory subsystems for multicore include increase in the number of levels of cache as well as increase in cache size. Traditional uni-core systems had a dedicated caching model whereas the recent multicore systems have a shared cache model. As a result of sharing and increase in the number of levels of cache, the cache access time increases and tends to consume a higher percentage of memory access time which in turn affects the execution time. A related issue which assumes significance is the effect of instruction set architectures on cache performance wherein different instruction set architectures influence the cache memory access time of programs.

In this paper we have tried to estimate the access time of shared cache memory on multi-core platforms for different ISA's. Results pertaining to execution time are also presented.

## 1.1. Related work

Substantial work is yet to be carried out on performance of cache memory systems for multicore platforms. Some of the previous work have been considered wherein:
1. In [4] by S. Laha et. al., Accurate low-cost methods for performance evaluation of cache memory systems have been discussed. In this work, results pertaining to trace driven simulation of predicting mean miss rate of cache have been presented.
2. Similarly, in [5] Cache performance of SPEC92 benchmark suite has been discussed in detail by J.E.Gee et.al. In this work issues pertaining to instruction cache miss ratio and data cache miss ratio have been discussed.
3. Our previous works have been on understanding the performance of multi-core platforms [6] and on the performance of cache on machines with X86 ISA [7]. Issues related to cache memory access and execution time of programs have been discussed using simulation tools such as DINERO IV and CACTI.
4. The work by A.C. Schneider et. al., [8] on dynamic optimization techniques deals with both ALPHA and X86 architectures on multi-core platforms.

We use similar techniques described in the above said work to compare the performance of ALPHA and X86 ISA.

## 1.2. Problem statement

The primary objective of this paper is the evaluation of the impact; caches have on different instructions set architectures. This is achieved by taking one particular benchmark from SPLASH2 and finding the access time on ALPHA ISA using M5sim and comparing the results with the results obtaining using CACTI on X86 ISA. The other issue that is addressed in this paper is the measure of execution time for varying sizes of cache. Also the impact of cache on execution time is demonstrated through simulation results using SPLASH-2 benchmark suite on M5sim. Here we have tried to find out whether benchmarks running on different instruction set architectures with a specific underlying hardware configuration produce results that are qualitatively similar.

The access time plays a key role in determining the execution time of any process. In a multi-core environment as the number of levels of cache increases, the cache access time tends to consume a major percentage of memory latency. Since the state of art multi-core systems have almost three levels of caches, it becomes essential to understand the impact of hierarchical cache design. This motivated us to study the influence of cache on both access time and execution time.

Section 2 describes the various tools used in the experiments followed by Section 3 describing the experimental setup and the parameter set used in the experiments. Section 4 discusses the summary of results followed by Section 5 dealing with future research and conclusion.

## 2. TOOLS USED

### M5SIM

M5 [9] is an emulation tool that is capable of performing event driven simulation. It enables users to simulate a multi-core environment with error modularity close to hardware. The represented model of system can be viewed as a collection of objects like CPU cores, caches, input and output devices. M5sim uses Object, Event, Mode driven architecture. Fig. 1 shows the basic block diagram of M5sim.

Objects are components such as CPU, memory, caches which can be accessed through interfaces. Events involve clock ticks that serve as interrupts for performing specific functionality. There are two basic modes of operation namely: full system mode and system call emulation mode. The major difference between the two modes of operation is that, the later executes system calls on a host machine whereas the former emulates a virtual hardware whose execution is close to the real system. In this paper, all the experiments are conducted using full system mode for alpha instruction set architecture. For further details regarding M5sim refer [10].

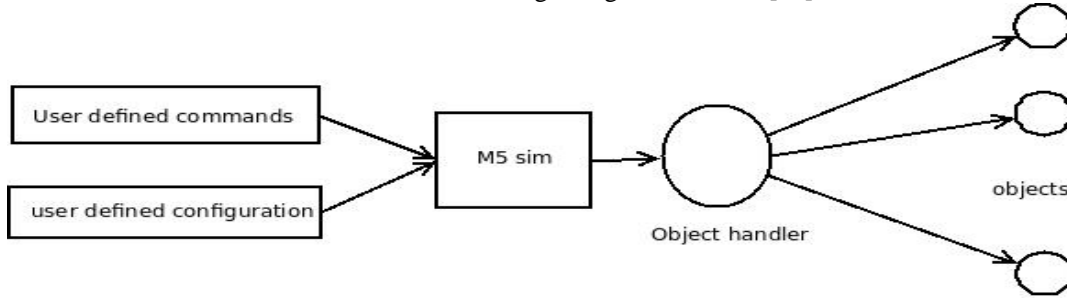

Figure 1.Block diagram of M5sim

### SPLASH-2 benchmark

SPLASH-2 [11] is a benchmark suite comprising of a set of applications making use of light weight threads and Argonne National Laboratories parmacs macro. With the help of this macro, SPLASH-2 invokes appropriate system calls instead of the standard fork system call. SPLASH-2 benchmark suite comes with benchmarks specific to applications and specific to kernel. In this paper, we produce the results obtained from the following three benchmarks namely radix sort, Fast Fourier Transformation (FFT) and Fast Multiple pole Method (FMM). The two former benchmarks come under kernel category and the later comes under application category. The benchmarks stated above are selected based on their relevance to multi-core processors and caches [12].

- Radix – This is an iterative algorithm, making use of all the processors. In a given iteration, a processor generates a key and passes over the key to the next processor. Once the key passes on, the processor generates a local histogram. The local histograms are combined to obtain a global histogram. Here, the steps involved in each iteration for generating the key requires all to all communication between processors.
- FFT – The input to this benchmark consists of n data points and n complex data points. The data points are modelled as n x n matrix. Each processor transposes a matrix of n/p x n/p, where p is the number of processors. This benchmark suite takes into consideration caching and blocks cache reuse for transpose of matrix.
- FMM – This simulates a system of bodies over a time-stamp. This benchmark uses unstructured communication among the system of bodies also referred to as cells. It computes the interactions among the cells and passes the effects.

### CACTI

CACTI [13] which refers to Cache Access and Cycle Time Indicator, is a cycle accurate simulator developed by Wada et.al. [14] for computing access and cycle time. Earlier work [15] by the authors on CACTI has given analytical insight into the aspects of access and cycle time. In this paper, some of the results of the previous work [7] have been considered to compute access time for comparison purposes. Here CACTI version 6.5 is used on an X86 instruction set architecture with number of cores parameter set to 2.

## 3. PROPOSED APPROACH

In this paper we have taken the uni-core model as proposed as in [16] and derived models for multi-core architecture using constructs similar to shared memory model.

### 3.1. Multicore shared cache model

In multi-core environment the cache is shared among multiple processors both at the core level and at the processor level. See Fig. 2, which shows sharing of L2-cache among multiple cores. The following list shows sharing of cache subsystems among processors.
- IBM power4 has 256 MB off-chip L3 cache,
- Itanium 2 and AMP Phenom II have 6 MB of on-chip L3 cache.
- Intel core i7 has 8 MB on-chip unified L3 cache.

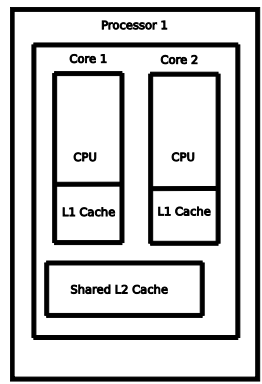

Figure 2. Multi-core shared cache model

In case of shared cache for multi-core model we mainly concentrate on average number of cache blocks that are accessed simultaneously. The entire modelling for shared cache can be divided into three different categories, namely:

**Caching involving only reads**

Consider a cache memory consisting of n blocks and let the total number of cores inside the processor be p. Let r be the number of requests generated by p cores. Here we consider that each processor core is capable of generating utmost one request.

$$\text{Number of cache blocks accessed simultaneously} = (1/n)^r$$

Since the operation performed is read, each request can be accessed simultaneously. Here, the probability of a cache block being accessed is (1/n). Since, the operation performed is read, the cache coherence problem does not arise.

**Caching involving only writes**

In case of a cache memory consisting of only write operation where the data from the main memory is updated onto the cache memory, multiple cores can update the value of single memory location or multiple cores can update two different memory locations.
Consider p cores, trying to update certain cache memory blocks. Four scenarios are described with respect to updating a cache memory block namely:

Case 1: cores writing the same content onto the same cache memory block,
Case 2: cores writing different content onto the same cache memory block,
Case 3: cores writing the same content onto different cache memory block and
Case 4: cores writing different content onto different cache memory blocks.

The mathematical model describing the above said conditions are described below.

$$P(case\ 1) = 1/n$$

$$P(case\ 2) = 1/n$$

$$P(case\ 3) = k/n$$

$$P(case\ 4) = k/n$$

Where k is the number of different values written by the processor-cores and $k < p$.

**Caching involving reads by certain cores and updates by certain other cores**

If the read and write operation are on different cache blocks then the number of operations performed is $1/n$. If they are performed on the same cache block, then the number of cache memory access performed is $k/n$, where k is the number of read or write operations performed, ($k<p$). p is the total number of cores. N is the total number of cache blocks. Based on the model described in [16] the CPU execution time is estimated as follows.

**CPU execution time estimation**

CPU execution time= [CPU clock cycles + Memory stall cycles$_{dedicated\ L1}$] * clock cycle time.

Memory stall cycle= Instruction count * Miss$_{dedicated\ L1}$/instruction * miss penalty$_{L1}$

Miss/instruction = Miss rate$_{dedicated\ L1}$+(Hit rate$_{shared\ L2}$ * Memory access/inst. + Miss rate$_{L2}$ *

Memory access/ inst * miss penalty$_{L2}$)

## 3.2. Existing processors

| No of cores | Code name | Brand | Family | Eg. | Technology (nm) | Clock speed GHz | L1 kB | L2 MB | L3 |
|---|---|---|---|---|---|---|---|---|---|
| 2 | Conroe | Xenon | 3xxx | | 65 | | 32 | 4 | |
| | Conroe-L | Celeron | 4x0 | 400 | 65 | .4 | 32 | 4 | |
| | Conroe-CL | Celeron | 4x5 | 445 | 65 | | 32 | 4 | |
| | Allendale | Conroe-2-duo | E4xxx | E4300 | 65 | 1.8 | 32 | 2 | |
| | | Xenon Core-2-duo | 3xxx | E4400 | 65 | 2 | 32 | 2 | |
| | | | | 3070 | 65 | 2.66 | 32 | 4 | |
| | | | | E6300 | 65 | 1.86 | 32 | 4 | |
| | | | | E6400 | 65 | 2.13 | 32 | 4 | |
| | | | | E6600 | 65 | 2.4 | 32 | 4 | |
| | | | | E6700 | 65 | 2.67 | 32 | 4 | |

| | | Celeron | E1xxx | E1600 | 65 | 2.4 | 32 | 512 kB | |
| | Wolfdale | Pentium Core-2-duo | E22xx | E2220 | 65 | 2.2 | 32 | 1 | |
| | | | E8xxx | E8000 | 45 | 2.66 | 32 | 6 | |
| | | Xenon | 31x0 | E3100 | 45 | 3.5 | 32 | 6 | |
| 4 | Kentsfield | Core 2 quad | Q6xxx | Q6600 | 45 | 2.4 | 32 | 2x4 | |
| | Yorkfield | Core 2 Extreme | QX6xxx | QX6700 | 45 | 2.67 | 32 | 2x4 | |
| | | Xenon | X33x0 | X3330 | 45 | 2.93 | 32 | 2x4 | |
| | | | X33x3 | X3333 | 45 | 3 | 32 | 2x3 | |
| | | Core 2 quad | QX8xxx | QX8100 | 45 | 3 | 32 | 2x6 | |

## 4. SIMULATION SETUP

The experiments are conducted on some important real-time applications such as fast fourier transformation, radix sorting and fast multiple pole method. These form a part of the SPLASH-2 benchmark suite. The parameters used in the experiments are mainly on varying the cache sizes. This paper tries to address the first order issue of the impact of cache on the performance of the multi-core system. The parameters in Table. 1 form the input set for execution of M5sim in full system mode. Though M5sim is capable of supporting a large number if system wide parameters, we take only the cache parameters into consideration and leave the rest of the parameters as default. For further details about the default parameters; refer [17]. In this paper we have considered dual-core and quad-core platform, along with L2 cache size of 8, 16, 32, 64 kB to understand the performance of cache on different instruction set architectures. The experiments on ALPHA ISA are conducted using M5sim and those on X86 are conducted using CACTI.

| Architecture | Cache for 65 nm and 45 nm technology | Cache size in MB | Associativity |
|---|---|---|---|
| ALPHA / X86 | L2 cache | 0.512 | 2 |
| ALPHA / X86 | L2 cache | 1 | 2 |
| ALPHA / X86 | L2 cache | 2 | 2 |
| ALPHA / X86 | L2 cache | 4 | 2 |
| ALPHA / X86 | L2 cache | 6 | 2 |
| ALPHA / X86 | L2 cache | 8 | 2 |
| ALPHA / X86 | L2 cache | 12 | 2 |

Table 1: List of architectures and cache size used in the experiments

## 5. RESULTS

The experiments are divided into two stages. The first set of experiments were conducted with execution time as y parameter and frequency of operation of core as x parameter. Here we have taken cache size of 64 kB. The second sets of experiments involve access time with respect to cache size.

The results were obtained by conducting the experiments on fairly equal number of instructions. From the graphs obtained in Fig. 3, we conclude that the data and instruction miss rate for L2 cache remains almost constant for each of the benchmarks taken into consideration. Fig. 3 shows that the instruction execution time remains qualitatively the same for varying frequency of operation. The number of instructions executed for Radix sort, fast fourier transformation and fast multiple pole method are 254089831, 254085595 and 254085421. From Fig. 3 we can conclude that, the frequency of execution of the multiple cores does not influence the execution time of the benchmarks considered in a qualitative manner.

Fig. 6 shows the number of hits during read operation, write miss rate and the average miss latency. The access time shown in Fig. 7 is calculated from the results in Fig. 6.

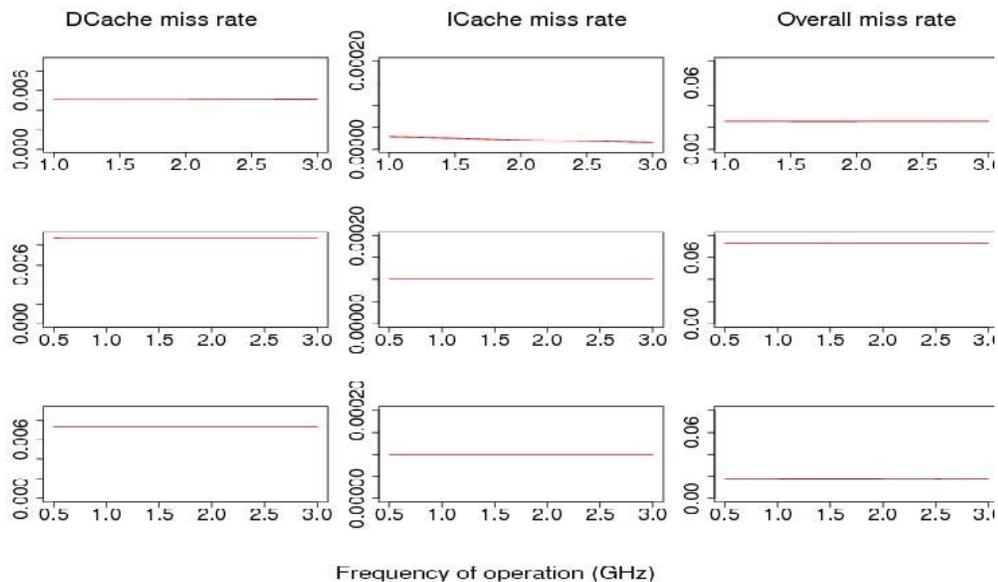

Figure 3: Miss rate for varying frequency of processor operation. The first row indicates the results for operations performed on Radix sort operation, the second row indicates fast fourier transform and the last row indicates the miss rate on fast multi-pole method.

The formula for average cache memory access time is obtained from the average memory access time as given in [16], where

$$\text{Access time} = \text{hit time} + (\text{miss rate} * \text{miss penalty})$$

Consider n1 be the number of hits during read operation, tn1 be the total number of instructions executed, e1 be the execution time in sim-seconds. Let write miss rate be denoted by wr1 and the average miss latency denoted as ml1. The access time is given by the relation:

$$\text{Average cache access time} = (n1 / tn1)*e1+(wr1 * ml1)$$

The access time in Fg. 7 is computed for ALPHA instruction set architecture using M5sim. Fig. 8 gives the access time computed on X86 architecture using CACTI. The results obtained on ALPHA and X86 architecture are qualitatively similar; see Fig. 7, 8.

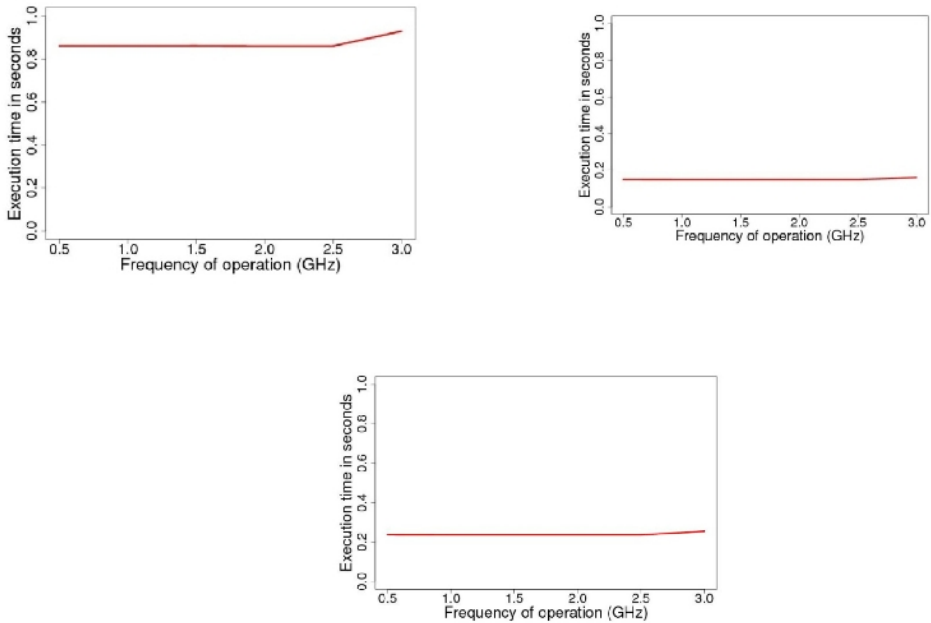

Figure 4. Execution time for radix sort, fast fourier transform and fast multiple pole method

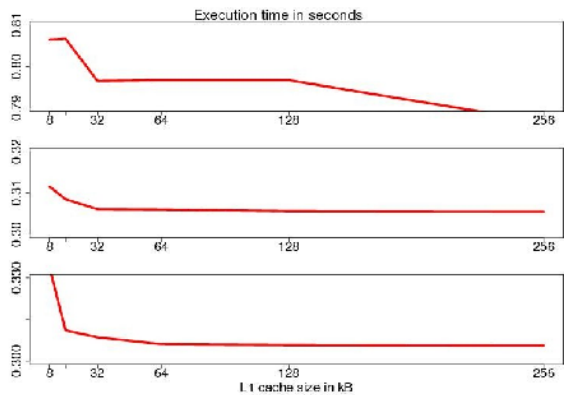

Figure 5. Cache size vs Execution time for Radix sort, Fast fourier transform and fast multiple pole method.

International Journal of Computer Science, Engineering and Applications (IJCSEA) Vol.1, No.5, October 2011

In this paper, we have described the impact of cache on execution time of processor on different instruction set architectures and an analysis of the processor frequency on execution time. We have studied the cache performance on multi-core architectures using simulators such as M5sim and CACTI on instruction set architectures like ALPHA and X86 with SPLASH-2 benchmark suite.

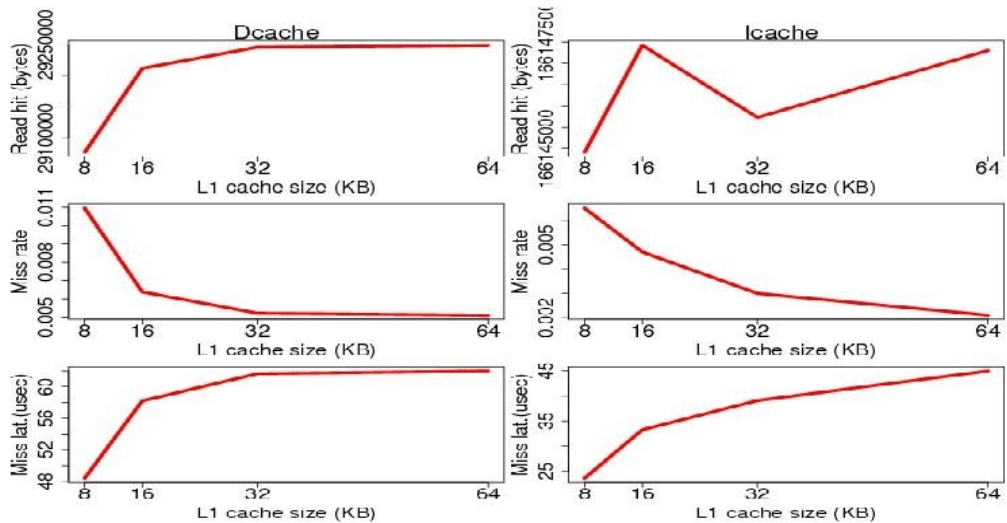

Figure 6. Read request hit, miss rate and miss latency (or) miss penalty for d-cache and I cache

The results obtained are for execution of 2541108434 instructions at a rate of 134633 instructions/second on M5sim on ALPHA ISA for fast fourier transformation. The comparison between results obtained from Fig. 7 of APLHA ISA and Fig. 8 of X86 architecture show that the access time are similar. We can infer from Fig. 7 and 8 that as the shared cache size increases, the access time increases. The access time reduces slightly as the number of cores increases. But, the comparison between the results of the two different ISA's show that, there is a significant improvement in the cache access time as we move from one ISA to another. Results of M5sim conducted on ALPHA ISA shows a reduced access time compared to the results obtained from CACTI on X86 ISA. The experiments were conducted for varying number of cores and we obtained similar results.

Results with 2 and 4 core imply that there may be slight decrease in access time with increase in number of cores. Table 2. Shows the improvement in execution time achieved with the reduction in cache access time. The impact of execution time will be significant with the increase in the number of instructions. Comparing the access time with ALPHA ISA and X86 ISA, the results indicate the possibility of reducing the access time with well-designed instruction set architectures.

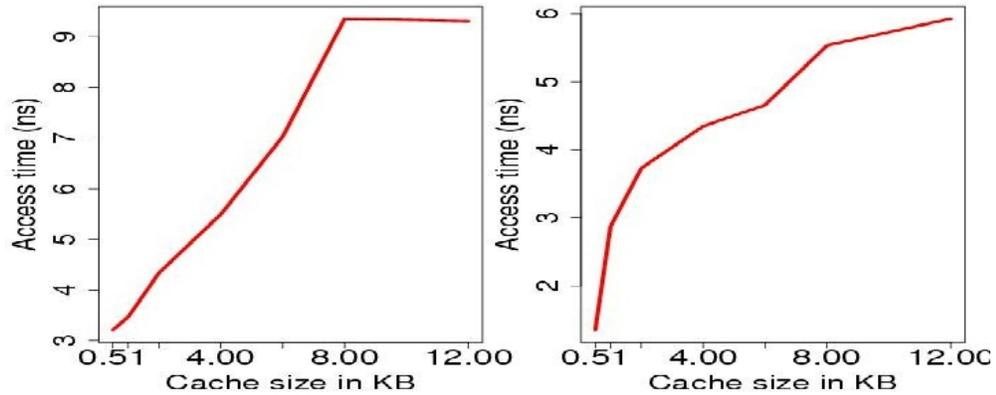

Figure 7. Access time calculated using M5sim for varying cache size on dual and quad core machines.

## 6. Avenues for future research

Future research would include an extensive analysis of all the results obtained as described in [6] [7] for a wide range of varying parameters. The previous work by the authors analysed the performance impact using tools and emulators only. An extension of the research would be to compare the results obtained using the simulators with that of real machines and to verify the correctness of results.

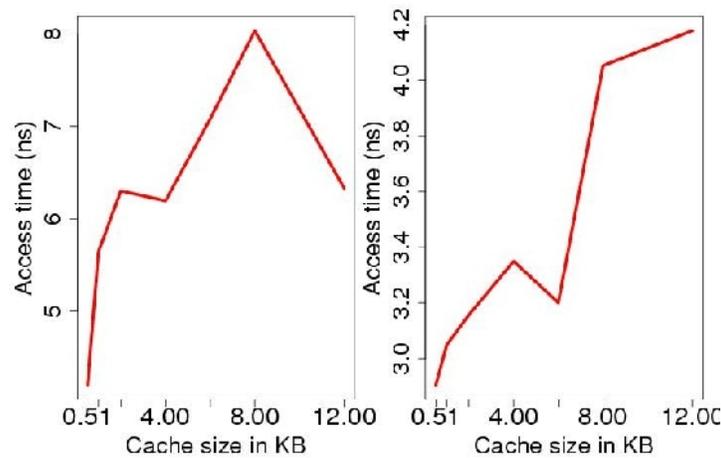

Figure 8. Access time calculated using CACTI for dual core and quad core platform.

| Cache size (MB) | 0.512 | 1 | 2 | 4 | 6 | 8 | 12 |
|---|---|---|---|---|---|---|---|
| Access time for dual core (ns) | 3.215 | 3.4634 | 4.3382 | 5.494 | 7.029 | 9.356 | 9.309 |
| Access time for quad core (ns) | 1.356 | 2.868 | 3.731 | 4.3497 | 4.66 | 5.534 | 5.928 |

Table 3: Access time calculated using M5sim for varying cache size on dual and quad core machines.

| Cache size (MB) | 0.512 | 1 | 2 | 4 | 6 | 8 | 12 |
|---|---|---|---|---|---|---|---|
| Access time for dual core (ns) | 4.2015 | 5.6446 | 6.3023 | 6.1949 | 7.0802 | 8.0356 | 6.3310 |
| Access time for quad core (ns) | 2.90356 | 3.04886 | 3.1573 | 3.3497 | 3.20066 | 4.0534 | 4.1792 |

Table 4: Access time calculated using CACTI for dual core and quad core platform.

| Core | Improvement in access time per inst. | No. of inst executed per second | Improvement in execution time (approx.). |
|---|---|---|---|
| ALPHA ISA | | | |
| 2-4 | 0.50625 ns | 134633 | 68.1579 ms |
| X86 ISA | | | |
| 2-4 | 0.2475 ns | 134633 | 33.32167 ms |

Table 5: Improvement in execution time with reduction in access time.

## 7. CONCLUSION

In this paper, the impact of cache memory subsystems on the execution time for different instruction set architectures has been discussed. In short, a study of cache performance on multicore architectures has been carried out using simulators such as M5sim and CACTI with ALPHA and X86 Instruction sets. The simulation results are encouraging and provide scope for further research in this area.

## ACKNOWLEDGEMENTS

The authors would like to thank the support received from the Intel Multi-core lab, Department of Computer Science and Engineering, National Institute of Technology, Tiruchirappalli for facilitating the research work.

## REFERENCES

[1] R. Kirner and P. Puschner, (2008) "Obstacles in worst-case execution time analysis", Proceedings of 11th IEEE International Symposium on Object Oriented Real-Time Distributed Computing, pp. 333-339.

[2] A.J. Smith, (2009) "Line (block) size choice for CPU cache memories", IEEE transaction on Computers, vol. 100, no. 9, pp. 1063-1075.

[3] K. Chakraborty, P.M. Wells, and G.S. Sohi, (2007) "A case for overprovisioned multi-core system: Energy efficient processing of multi-threaded programs", Technical Report of CS-TR-2007-2607, University of Wisconsin-Madison

[4] S. Laha, J.H. Patel and R.K. Iyer, (2002), "Accurate low-cost methods for performance evaluation of cache memory systems", IEEE Transactions on Computers, vol. 11, pp. 1325-1336.

[5] J.D. Gee, M.D. Hill, D.N. Pnevmatikatos, and A.J. Smith, (2002), "Cache performance of SPEC92 benchmark suite", Proceedings of IEEE Micro, vol. 13, pp. 17-27.

[6] V.V. Srinivas and N. Ramasubramanian, (2011), "U?nderstanding the performance of multi-core platforms", International Conference on Communications, Network and Computing, CCIS-142, pp. 22-26.

[7] N. Ramasubramanian, V.V. Srinivas and P. Pavan kumar, (2011), "Understanding the impact of cache performance on multi-core architectures", International Conference on Advances in Information Technology and Mobile Communication, CCIS-147, pp. 403-406.

[8] A.C. Schneider Beck Fl, L. Carro, (2010), ""Dynamic Optimization Techniques", Journal on Dynamic Reconfigurable Architectures and Transparent Optimization techniques, pp. 95-117.

[9] S.K. Reinahardt, N. Binkert, A. Saidii and R.D.K. Lim, (2005), "Understandint the M5 simulator", ICSA tutorials and workshop.

[10] M5 simulator, http://www.m5sim .org

[11] C. Bienia, S. Kumar and K. Li, (2008), "Parsec vs SPLASH2: A quantitative comparison of two multithreaded benchmark suites on chip multi-processors", IEEE International Symposium on Workload Characterization, pp. 47-56.

[12] Steven Cameron Woo, Moriyoshi Ohara, Evan Torrie, Jaswinder Pal Singh and Anoop Gupta, (1995), "The SPLASH-2 programs: Characterization and methodological considerations", proceedings of 22nd Annual International Symposium on Computer Architecture, pp. 24-36.

[13]  CACTI-6.5, http://www.hpl.hp.com/research/cacti/.

[14]  T. Wada, S. Rajan and S.A. Przybylski, (2002), "An analytical access time model for on-chip cache memories", IEEE Journal of Solid Ctate Circuits, vol. 27, pp. 1147-1156.

[15]  N. Muralimanoharan, R. Balasubramonian and N.P. Jouppi, (2008), "Architecting efficient interconnects for large cache with CACTI 6.0", Proceedings of IEEE Micro, vol. 28, no. 1, pp. 69-79.

[16]  J.L. Hennessy and D.A. Patterson, (2007), "Computer Architecture: A Quantitative Approach", 4$^{th}$ edition, Elsevier Inc.

[17]  Lei, Shi, P. Jun, Y. Lei, Z. Tiejun and W. Donghui, (2009), "Kernel Analysis and Application of M5 Simulator", Journal of Microcomputer Applications, vol. 4.

**Authors**

N. Ramasubramanian is an Associate Professor in the Department of Computer Science and Engineering at National Institute of Technology – Tiruchirappalli, India. His research interests are multi-core architecture, advanced digital design and distributed systems.

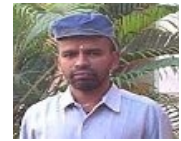

Srinivas V.V. is a graduate student in the Department of Computer Science and Engineering at National Institute of Technology-Tiruchirappalli, India. His areas of research include multi-core computing, parallel programming and networking.

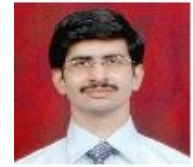

Dr. N. Ammasai Gounden is a Professor in the Department of Electrical and Electronics Engineering at National Institute of Technology – Tiruchirappalli, India. His areas of research include power electronic controllers for wind, solar applications and Microprocessor and Microcontroller based applications.

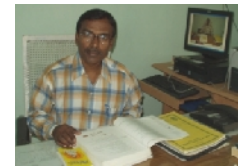